\documentstyle[twocolumn,aps,floats]{revtex}

\begin{document}

\draft \tolerance = 10000

\setcounter{topnumber}{1}
\renewcommand{\topfraction}{0.9}
\renewcommand{\textfraction}{0.1}
\renewcommand{\floatpagefraction}{0.9}

\twocolumn[\hsize\textwidth\columnwidth\hsize\csname
@twocolumnfalse\endcsname

\title{Physical Consequences of Moving Faster \\ than Light in Empty Space}
\author{L.Ya.Kobelev   \\
Department of  Physics, Urals State University \\
 Av. Lenina, 51, Ekaterinburg 620083, Russia \\
 E-mail:  leonid.kobelev@usu.ru   }
\maketitle

\begin{abstract}
Physical phenomena caused by particle's moving faster than light
in a space with multifractal time with dimension close to integer
($d_{t}=1+\varepsilon({\bf r}(t),t),\; |\varepsilon| \ll 1$ - time
is almost homogeneous and  almost isotropic) are considered. The
presence of gravitational field is taken into account. According
to the results of the developed by the author theory, a particle
with the rest energy $E_{0}$ would achieve the velocity of light
if given the energy of about $E\sim 10^{3}E_{0}$.
\end{abstract}

 \vspace{1cm}

]

\section{Introduction}
In \cite{kob1}, basing on the theory of multifractal time and
space proposed in \cite{kob2}, the main features of relative
motion of systems close to inertial ("almost" inertial systems) in
the space with fractal dimension of time $d_{t}=d_{t}({\mathbf
r},t)=1+\varepsilon, \, |\varepsilon|\ll1$ were formulated, and it
was shown that in such systems motion of body with any velocity
(from zero up to infinite) becomes possible. For total momentum
and energy of a moving object the following relations were
obtained
\begin{equation}  \label{1}
p=\beta^{*^{-1}}m_{0}v=\frac{m_{0}v}{\sqrt[4]{\beta^{4}+4a_{0}^2}}
,\,\,\,
E=E_{0}\sqrt{\frac{{v^{2}c^{-2}}}{\sqrt{\beta^{4}+4a_{0}^2}}+1}
\end{equation}
where  $\beta=\left|1-v^{2}/c^{2}\right|^{1/2},\,
a_{0}=\sum_{i}\beta_{i}{\bf F}_{0i}\frac{{\bf v}}{c}ct, \, {\bf
F}_{i}=dL_{i}/d{\bf r}$ with $L_{i}$ standing for Lagrangian
density of $i$th physical field, $t$ for time and $\beta_{i}$
being dimensional factors for $i$th field, providing
dimensionlessness of $\varepsilon$:
$\varepsilon=\sum_{i}\beta_{i}L_{i}$. However, the question on
what new physical phenomena can be observed from the point of view
of this theory if a body's velocity exceeds that of light in empty
space was temporarily put aside. The present paper is devoted to
fill in this gap and deals with several consequences of the
proposed multifractal concept of time \cite{kob1} and relations
(\ref{1}), that allow for experimental verification.

\section{Vavilov-Cherenkov-like radiation at $v>c$}

According to the main statements of \cite{kob1}, small
noninertiality of moving frames of reference arising from the time
being multifractal, and openness of any real system (for
statistical theory of open systems see \cite{klim}) must lead to
small deviations from the conservation laws. In particular, the
law of energy conservation would be fulfilled only approximately.
Basing on the laws of electrodynamics, which remain valid in the
theory of multifractal time at any velocities, it was shown that
when a body's velocity $v$ reaches that of light $c$ and continue
increasing, the energy of the moving body reaches it maximum value
and then begins diminishing (see (\ref{1}). As this takes place,
more and more energy being lost by the body in order to span the
small deviation from the energy conservation will be emitted to
the surrounding matter (space) through radiation, quite similar to
the Vavilov-Cherenkov radiation. But, unlike the tachion theory,
here the motion faster then light is done by ordinary particles.
Moreover, free motion then is always a motion with acceleration,
for it is velocity growth that is accompanied by energy decreasing
in this region (we still consider the principle of energy minimum
in equilibrium state to be valid). In the region of velocities
greater than $c$, a source of energy is needed in order to {\em
decelerate} a moving particle! Thus, increasing the velocity
beyond the speed of light leads to energy loss accompanied by
radiation. With the energy of this radiation, the energy
conservation law is "almost" fulfilled
\begin{equation}
E=E_{0}\sqrt{\frac{v^{2}}{c^{2}\beta^{*^{2}}}+1}+E_{rad} \;\;\;\;
v>c
\end{equation}
It can be shown that at the velocities greater than the speed of
light perturbation of multifractal structure of time still remains
negligibly small, just as it is the case for resting particles.

\section{Does the casuality remain in $v>c$ region?}

In special relativity the necessary condition of two events that
take place in a fixed frame of reference in points $x_{1}$ and
$x_{2}$ at times $t_{1}$ and $t_{2}$ and of the same events in a
moving frame to be casually connected is the validity of the
following unequalities
\begin{equation}
t_{2}-t_{1}>0,\,\,\, t_{2}^{\prime}-t_{1}^{\prime}=
\frac{t_{2}-t_{1}}{\sqrt[4]{\beta^{4}+4a_{0}^2}}
\left(1-\frac{v}{c^{2}}v_{inf}\right)>0
\end{equation}
where $v_{inf}$ is the velocity of the influence spreading between
the points $x_{1}$ and $x_{2}$. If $v_{inf}<c^{2}/v$, the casual
connectivity of the two events remains at any speed of motion $v
\le \infty$. Though, when $v=\infty$, the rate of influence must
be zero. Nevertheless, if this unequality is not fulfilled
($v_{inf}v>c^{2}$), casuality does not remain. This violation of
casuality for different events is one of the main arguments
against the tachion theory, restricting, in particular, the free
will of objects. Fortunately, in our model casuality does not
break for the following reason. Any body moving faster than light
radiates energy, and the greater its speed the less its energy.
Hence, motion with velocity greater than that of light is always a
motion with acceleration and thus is not "almost inertial" in our
terminology. Therefore, in terms of the assumptions made, any
comparison between moving with acceleration and fixed frames of
reference have no physical sense.

Stress now the main differences between motion faster than light
in tachion and in multifractal time models. In the tachion model
the whole region of velocities consists of two separate parts,
velocities greater and less than that of light in vacuum.
Particles whose speed is greater than $c$ (tachions) can not cross
the barrier and move into the other region. Tachions have several
peculiarities in their motion, and the principle of casuality can
be violated if we are to compare events in different regions of
velocities. In the model of multifractional time any particle, if
supplied with sufficiently big energy, can be accelerated up to
velocities greater than the speed of light, and thus found itself
in the tachion region. However, in this region it will be
constantly radiating energy and moving with growing velocity. In
this process its energy tends to a finite value $E\to
E_{0}\sqrt{2}$ as $v \to \infty$. By consuming energy it, though,
can be slowed down, and, having its energy being increased up to
$E=E_{0}/\sqrt{2a_{0}}$ (see (\ref{1})), can return in the region
$v<c$ and begin to radiate again, but now when being decelerated.

\section{Possible physical phenomena at $v>c$}

A body (with nonzero rest mass) moving with the speed of light has
maximum possible energy and represents a sort of energy reservoir
- if its velocity increases or decreases, the excess of energy
emits through radiation. Such a body thus can serve as an energy
source, since small initial (e.g., spontaneous) increasing of its
velocity would lead to release of immense energy of order $E\sim
E_{0}\sqrt{a_{0}}$. In this connection, the following possible
consequences allowing for experimental observation and
applications of motion faster then light can be pointed out.

a) a sudden burst of radiation can occur as a particle's velocity
increases from $v=c$ to $v>c$. As an example of observation of
this effect we can consider a charged particle in accelerators
like synchrotron. At energies much greater then the rest energy
($v\approx c, \,v<c$), the particle's velocity almost does not
alter while its energy can vary by orders of magnitude, with this
change accompanied by considerable growth of synchrotron
radiation. When velocity reaches the value $v=c$, energy has its
maximal value (see (\ref{1})). Then in the narrow region of
velocities $0<v^{2}/c^{2}-1\le 4a_{0}$ the particle looses almost
all its energy through Vavilov-Cherenkov-like radiation and
synchrotron radiation. In this process the radius of its orbit
remains almost the same (the particle's velocity is still close
$c$). As this occurs, the radiation power grows sharply and has
the order of $E_{0}10^{3}t^{-1/2}sec^{1/2}$ (which equals to $\sim
10^{12}t^{-1/2}eV sec^{1/2}$ for protons. This jump of radiation
power can be detected by registering appearance of high-energy
$\gamma$-quantums, mesons, electron-positron pairs etc. Further
increasing of velocity will result in the particle's getting out
from the stationary orbit and becoming invisible for the observer.
The latter is connected with the fact that in order to reduce
particle's velocity down to $v=c$ it is necessary to give it
energy it lose through radiation ($\sim 10^{12}eV$ during one
second for proton). Such particle will move undergoing
acceleration without substantial change in energy
($E_{min}=E_{0}\sqrt{2}$). Experimental observation then becomes
possible only if it collides with something that can supply it
with the required for the transition in the region $v<c$ amount of
energy (another high energy particle or $\gamma$-quantum). In this
case the particle can be detected as ordinary charged particle
with very high energy, that ionizes matter and gives birth to
bunches of $\gamma$-quantums, mesons, electron-positron pairs etc.

b) Propagation of a faster-than-light particles beam in a dense
media would lead to diminishing the media's temperature due to
such particles' attaining energy while being decelerated, and this
can serve as a possible method to decrease the energy of a hot
dense matter (thermonuclear plasma, neutron star, nuclear power
reactors etc.) This method of cooling very hot matter with high
density and high scattering crossection for faster-than-light
particles may turn to be one of the most effective ways of doing
that for these kinds of media because of huge amounts of energy
necessary to slow down such particles.

c) Energy $E_{0}10^{3}t^{-1/2}sec^{1/2}$ released over small time
intervals corresponds to the temperature of $T\sim
E_{0}10^{19}t^{-1/2}sec^{1/2}K$, and perhaps can be used for
initiating or controlling thermonuclear fusion in
deuterium-tritium media.

d) Provided that the conditions for coherent radiation appearance
are satisfied, a laser with working frequency of $\nu \sim
10^{3}E_{0}\hbar^{-1}t^{-1/2}sec^{1/2}$ can probably be created.

\section{Conclusion}

Investigation of relative motion in the multifractal time model
for "almost inertial" systems \cite{kob1} indicates of possible
appearance of a number of new physical phenomena in the region of
the velocities greater than the speed of light and new notions
about properties of mass and energy as functions of velocity in
the case of fractional corrections to the topological dimension of
time being small (in particular, possibility of motion with any
velocity, existence of a maximum energy for a particle and several
others). The proposed model, based on the concept of time with
fractional dimension, does not contradict to special relativity
and is not its generalization. Indeed, all motions and frames of
reference in this model are absolute. Due to the time and space
inhomogeneity and openness of time-space in general, Galileee and
Lorentz transformations, and conservation laws are only
approximate. This does not contradict to the usually observed
phenomena, since the deviations from the classical laws are very
small (as it is the case with space-time curvature in general
relativity). The model of relative motion in multifractal time, on
which the present paper is based, hence represents a theory of
relative motion in "almost inertial" frames of reference in the
space of almost homogeneous time with dimension very close to
integer. The theory proposed in \cite{kob1} contains the special
relativity as a special case, corresponding to zero fractional
corrections to the dimension of time, and reduces to it if we set
time dimension to be unity (then all the approximate laws named in
the paper become exact). Approximate validity of the Lorentz
transformations follows from the assumption about the smallness of
fractional correction to the topological dimensionality of time.
On the other hand, one should not be surprised by taking into
account the velocity of light dependence on the Lagrangian
densities of physical fields in "almost inertial" frames. For
example, in general relativity it depends on gravitational
potentials. As it was mentioned in \cite{kob2}, appearance of
fractional dimensions of space and time can be interpreted in
terms of Penrose's ideas \cite{pen} concerning appearance of the
equations of free physical fields as a result of deformations of
certain complex manifolds (such as co-homologies of bunches with
coefficients) characterizing space-time (in our case, with
fractional dimension)

The main results of the model of multifractal time that disappear
when we use usual concept of time are the following.

1. The possibility for any object to move faster than light
(instead of the factor $\beta=\sqrt{1-v^{2}/c^{2}}$ of special
relativity  the modified factor $\beta^{*}=\sqrt[4]{\beta^{4} +
4a_{0}^{2}}$ appears)

2. Total energy at $v>c$ is determined by the expression
$$E=\sqrt{p^{2}c^{2}+E_{0}^{2}}=
E_{0}\sqrt{\frac{1+\beta^{2}}{\sqrt{\beta^{4} + 4a_{0}^{2}}}+1}
,\,\,\, E_{0}=m_{0}c^{2}$$ with $\beta^{2}=\frac{v^{2}}{c^{2}}-1$
which does not coincide with the relation $E=\beta^{*^{-1}}E_{0}$
valid for $v<c$

3. Maximal value of the total energy (mass, momentum) are bounded
by the value corresponding to the motion with the velocity of
light $$E_{max}=m_{max}c^{2}=E_{0}\sqrt{2a_{0}}, \,\,\,
p_{max}=m_{max}c$$ Both energy and momentum remain finite as $v\to
\infty$ $$E_{\infty}=E_{0}\sqrt{2},\,\,\, p_{\infty}=m_{0}c $$

4. Existence of Vavilov-Cherenkov-like radiation not connected
with deceleration processes

5. If the fractional correction $\varepsilon$ to the time
dimension is zero, our model fully reduces to the equations and
conclusions of special relativity.

The energies, necessary according to our theory to accelerate the
particles up to the velocity of light ($E\sim 10^{10}$eV for
electron, $E\sim 10^{12}$eV for proton) seem to be available in
the nearest decade, thus making  the experimental verification of
the theory of "almost inertial" frames of reference \cite{kob1}
possible.

\end{document}